\documentclass[preprint,prx,longbibliography]{revtex4-1}
\usepackage{subcaption}
\usepackage{graphicx}
\usepackage{comment}
\usepackage{amsmath,amssymb,amsthm} 
\usepackage{scalerel,stackengine}
\usepackage{bm}
\usepackage{verbatim}
\usepackage{float}

\begin{document}
	\title{  Newton's algorithm for discrete classical dynamics }

\author{ S\o ren  Toxvaerd }
\affiliation{ Department of Science and Environment, Roskilde University, Post box 260, DK-4000 Roskilde, Denmark}
\email{st@ruc.dk; to appear in J. Chem. Phys.}
\begin{abstract}
A recent article in J. Chem. Phys. argues that the two algorithms,
the velocity-Verlet,  and position-Verlet integrators, commonly used in Molecular Dynamics (MD) simulations,  are different \cite{Ni2024}. But
not only are the two algorithms just different formulations of the same discrete algorithm,
	but so are other simple discrete algorithms used in MD
in the natural sciences. They are all reformulations of the discrete algorithm derived by Newton in 1687 in $\textit{Proposition I}$ in the
very first part of his book $Principia$. The different reformulations of Newton's algorithm for discrete dynamics lead to identical discrete dynamics
with the same invariances, momentum, angular momentum, and energy as Newton's analytical dynamics. 
Hundreds of thousands of MD simulations with Newton's discrete 
dynamics have appeared,
 but unfortunately with many recorded errors for energies, potential energies, temperatures, and heat capacities. The public software for MD should be
corrected.
\end{abstract}
\maketitle

\section{Introduction}
A recent article in J. Chem. Phys. argues that the two simple discrete algorithms, the velocity-Verlet,
and position-Verlet integrators, are different \cite{Ni2024}. But
not only are the two algorithms just different formulations of the same discrete algorithm, 
so are other simple discrete algorithms used in Molecular Dynamics (MD) simulations
in the natural sciences. They are all reformulations of the discrete algorithm derived by Newton in 1687 in $\textit{Proposition I}$ in the
very first part of his book $Principia$. 
 \cite{Newton1687,Toxvaerd2023}.

The new position of an object in Newton's  discrete dynamics, $\textbf{r}_i(t+\delta t)$, at time $t+\delta t$ 
$i$ with the mass $m_i$   is determined by
the force $\textbf{f}_i(t)$ acting on the object   at the discrete position $\textbf{r}_i(t)$  at time $t$ together with 
 the position $\textbf{r}_i(t-\delta t)$ at $t - \delta t$  as
\begin{equation}
	 m_i\frac{\textbf{r}_i(t+\delta t)-\textbf{r}_i(t)}{\delta t}
			=m_i\frac{\textbf{r}_i(t)-\textbf{r}_i(t-\delta t)}{\delta t} +\delta t \textbf{f}_i(t),	
 \end{equation}
where the velocities  $ \textbf{v}_i(t+\delta t/2) =  (\textbf{r}_i(t+\delta t)-\textbf{r}_i(t))/\delta t$ and
 $  \textbf{v}_i(t-\delta t/2)=  (\textbf{r}_i(t)-\textbf{r}_i(t-\delta t))/\delta t$ and corresponding momenta are constant in
the time intervals in between the discrete positions.

\begin{figure}
\centering
\begin{subfigure}[b]{1.00\linewidth}
\includegraphics[width=6.6cm]{{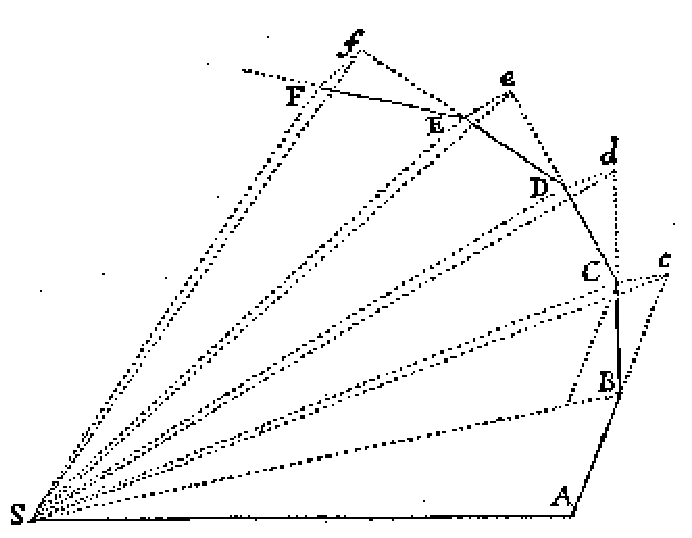}}
\caption{A Newton's figure at $Proposition$ $I$ in Principia, with the formulations of the discrete dynamics. 
}\label{fig:Proposition I}
\end{subfigure}

\begin{subfigure}[b]{1.00\linewidth}
\includegraphics[width=5cm,angle=-90]{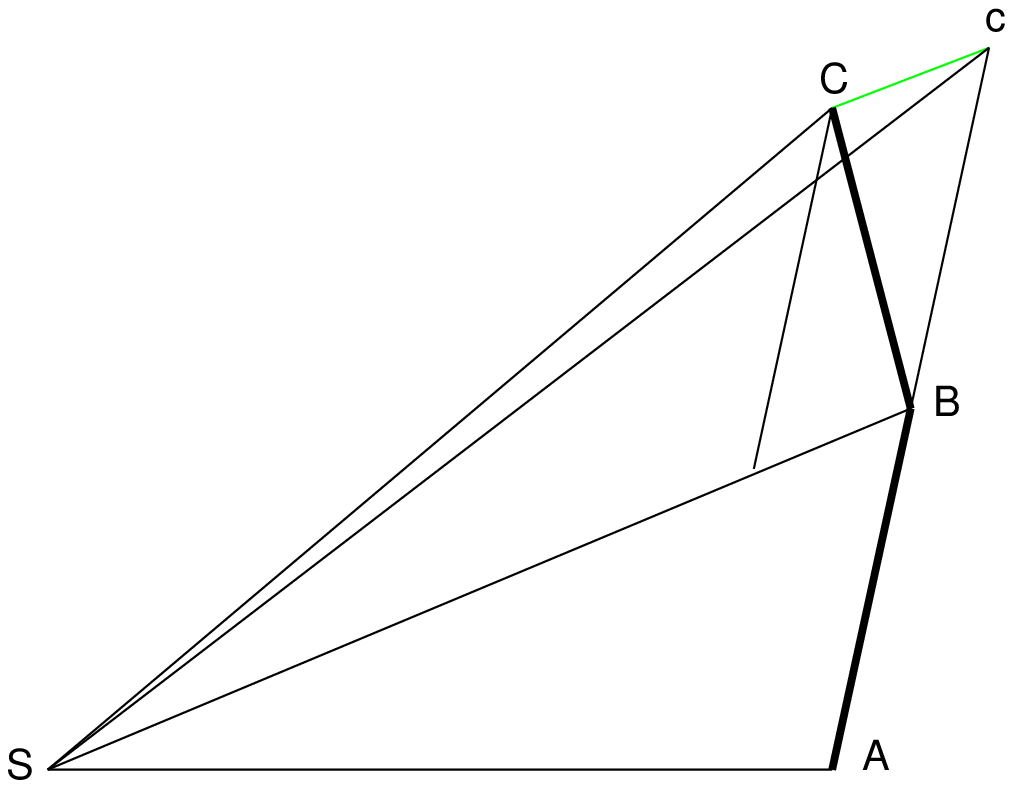}
	\caption{ Central part: A: $\textbf{r}_i(t-\delta t)$;  B:  $\textbf{r}_i(t)$;  C: $\textbf{r}_i(t+\delta t)$, etc.. The deviation cC (green)
 	  from the straight line ABc (Newton's first law) is caused by the 
	  force $\textbf{f}_i(t)$ with direction $\overrightarrow{\textrm{B}\textrm{S}}$ at time $t$.	} 
\end{subfigure}
\label{fig:Scematic}
	\caption{ Newton's figure at $Proposition$ $I$} 
\end{figure}

\section{Newton's discrete algorithm}
Newton  begins \textit{Principia} by postulating Eq. (1) in \textit{Proposition I} and with Figure 1. The English translation
of \textit{Proposition I} is

\textit{Of the Invention of Centripetal Forces.}\\
  PROPOSITION I. Theorem I.\\
 \textit{The areas, which revolving bodies describe by radii drawn to an immovable centre of force do lie in the same immovable planes, and
 are proportional to the times in which they are described}.\\
\textit{ For suppose the time to be divided into equal parts, and in the first part of time let the body by its innate force describe the right line
AB. In the second part of that time, the same would (by Law I.), if not hindered, proceed directly to c, along the line Bc equal to AB; so that the radii
AS, BS, cS, drawn to the centre, the equal areas ASB, BSc, would be described. But when the body is arrived at B,
\textbf{suppose that a centripetal force acts at once
with a great impulse}, and, turning aside the body from the right line Bc, compels it afterwards
to continue its motion  along the right line BC. Draw cC parallel
to BS meeting BC in C; and at the end of  the second part of the time, the body (by Cor. I of Laws) will be found in C, in the same plane with the
triangle ASB. Join SC, and,  because SB and Cs are parallel, the triangle SBC will be equal to the triangle SBc, and therefore also to the
triangle SAB. By the like argument, if the centripetal force acts successively in C, D, E, \& c., and makes the body,
in each single particle of time, to describe the right lines CD, DE, EF, \& c., they will all lie in the same plane;
and the triangle SCD will be equal to the triangle SBC, and SDE to SCD, and SEF to SDE. And therefore, in equal times, equal areas are described in on immovable plane:
and, by composition, any sums SADS, SAFS, of those areas, are one to the other as the times in which they are described. Now let the number of
those triangles be augmented; and their breadth diminished in infinitum; and (by Cor. 4, Lem III) their ultimate perimeter ADF will be a curve line:
and therefore the centripetal force, by which the body is perpetually drawn back from the tangent of this curve, will act continually; and any described
areas SADS, SAFS, which are always proportional to the times of description, will, in this case also, be proportional to those times.} Q. E. D.

The central assumption ..\textit{\textbf{suppose that a centripetal force acts at once with a great impulse}}
in \textit{Proposition I}  is highlighted here. The forces change the momenta only at discrete times, 
and the dynamics is solely determined by the positions and the forces at
these discrete times.
The   positions  with constant velocities and momenta in between the discrete times
are changed with $\overrightarrow{\textrm{A}\textrm{B}}= \textbf{r}_i(t)-\textbf{r}_i(t-\delta t)=\overrightarrow{\textrm{B}\textrm{c}}$ (Newton's first law)
to  $\overrightarrow{\textrm{B}\textrm{C}}= \textbf{r}_i(t+\delta t)-\textbf{r}_i(t)$. The change in position is caused by the
force $\textbf{f}_i(t)$ in the direction  $\overrightarrow{\textrm{B}\textrm{S}}$, which
within the next times to $t+\delta t$, change the momentum with a total amount $\delta t\textbf{f}_i(t)$ and the  position  to C with
$\overrightarrow{\textrm{c}\textrm{C}}= \delta t^2 \textbf{f}_i(t)/m_i$, i.e.
\begin{equation}
	\overrightarrow{\textrm{B}\textrm{C}}=\overrightarrow{\textrm{A}\textrm{B}}+ \overrightarrow{\textrm{c}\textrm{C}},
 \end{equation}
 with is equal to Eq. (1).  Ref. \cite{Toxvaerd2023} is  a review of Newton's discrete dynamics.

 Newton obtained his second law for classical analytic dynamics as the limit $ lim_{\delta t \rightarrow 0}$ of Eq. (1).
 Today Newton's second law is formulated as an equality between the mass times the acceleration being
 equal to the force acting on the object, but this formulation is due to Euler in 1736 after Newton died in 1727 \cite{Coelho2018}.

\section{Reformulations of Newton's discrete algorithm}

  The algorithm, Eq. (1), is usually  presented  as the Leapfrog algorithm 
\begin{eqnarray}
	\textbf{v}_i(t+\delta t/2)=  \textbf{v}_i(t-\delta t/2)+ \frac{\delta t}{m_i}  \textbf{f}_i(t) \nonumber \\
\textbf{r}_i(t+\delta t)= \textbf{r}_i(t)+ \delta t \textbf{v}_i(t+\delta t/2),	  
\end{eqnarray}	  
where the new values $\textbf{v}_i(t+\delta t/2)$
and $\textbf{r}_i(t+\delta t)$ are obtained from the
corresponding old values  $\textbf{v}_i(t-\delta t/2)$ and $\textbf{r}_i(t)$ .
The rearrangement of Eq. (1) gives  the Verlet algorithm \cite{Verlet1967,Levesque2018}
\begin{equation}
	\textbf{r}_i(t+\delta t)=2\textbf{r}_i(t)-\textbf{r}_i(t-\delta t) +\frac{\delta t^2}{m_i}\textbf{f}_i(t).	  
\end{equation}	

The velocity-Verlet algorithm \cite{Swope1982}
\begin{eqnarray}
	\textbf{r}_i(t+\delta t)=  \textbf{r}_i(t)+  \delta t \textbf{v}_i(t)+ \frac{\delta t^2}{2m_i}  \textbf{f}_i(t)  \\
	\textbf{v}_i(t+\delta t)= \textbf{v}_i(t)+ \frac{\delta t}{2m_i}[\textbf{f}_i(t) +  \textbf{f}_i(t+\delta t)],	  
\end{eqnarray}	 
with 
\begin{equation}
	\textbf{v}_i(t)=\frac{\textbf{v}_i(t+\delta t/2)+\textbf{v}_i(t-\delta t/2)}{2}=  \frac{\textbf{r}_i(t+\delta t) -\textbf{r}_i(t-\delta t)}{2\delta t}
\end{equation}
is a reformulation of Newton's discrete algorithm. It can be seen by rearranging Eq. (5)
\begin{equation}
	\textbf{v}_i(t)=\frac{\textbf{r}_i(t+\delta t)-\textbf{r}_i(t)}{\delta t}- \frac{\delta t}{2m_i } \textbf{f}_i(t) 
\end{equation}
and inserting Eq. (8) in  Eq. (6)
\begin{equation}
	\textbf{v}_i(t+\delta t)= \frac{\textbf{r}_i(t+\delta t)- \textbf{r}_i(t)}{\delta t} +\frac{ \delta t}{2m_i} \textbf{f}_i(t+\delta t),	
\end{equation}
or
\begin{equation}
 \frac{\textbf{r}_i(t+2\delta t) -\textbf{r}_i(t)}{2\delta t}
	= \frac{\textbf{r}_i(t+\delta t)- \textbf{r}_i(t)}{\delta t} +\frac{ \delta t}{2m_i} \textbf{f}_i(t+\delta t),	
\end{equation}
which by a rearrangement is the Verlet algorithm
\begin{equation}
 \textbf{r}_i(t+2\delta t)
	= 2\textbf{r}_i(t+\delta t)- \textbf{r}_i(t) + \frac{\delta t^2}{m_i} \textbf{f}_i(t+\delta t).	
\end{equation}

The position-Verlet algorithm, Eq. (2.22) in Ref. \cite{Tuckerman1992}, is
\begin{eqnarray}
	\textbf{v}_i(t+\delta t)= \textbf{v}_i(t)+ \frac{\delta t}{m_i}  \textbf{f}_i( \textbf{r}_i(t+\delta t/2)) \\
	\textbf{r}_i(t+\delta t)=  \textbf{r}_i(t)+  \frac{\delta t}{2}[ \textbf{v}_i(t)+  \textbf{v}_i(t+\delta t)], 
\end{eqnarray}	 
and the algorithm differs from the other discrete algorithms  by that the forces are calculated at the positions
\begin{equation}
	\textbf{r}_i(t+\delta t/2)=  \textbf{r}_i(t)+  \frac{\delta t}{2} \textbf{v}_i(t)
\end{equation}
 after a half timestep
in between the positions  $\textbf{r}_i(t)$ and  $\textbf{r}_i(t+\delta t)$. 
However,
Newton's discrete dynamics depends solely on the momenta $m_i \textbf{v}_i(\tilde{t})$ at the time $\tilde{t}$ where the forces act, so we need  to 
compare different  algorithms with  changes in velocities and  momenta  at the time where they change, and to change the time when one
compares the position-Verlet algorithm with the other reformulations of Newton's discrete algorithm. Doing so
\begin{equation}
	\tilde{t} \equiv t +\delta t/2,
\end{equation}
and
\begin{eqnarray}
	\textbf{r}_i(\tilde{t})=\textbf{r}_i(t+\delta t/2)\\
  \textbf{v}_i(t)=\textbf{v}_i(\tilde{t}-\delta t/2), 
\end{eqnarray}
and Eq. (14) is
\begin{equation}
	\textbf{r}_i(\tilde{t})=  \textbf{r}_i(t)+  \frac{\delta t}{2} \textbf{v}_i(t).
\end{equation}
The change of velocities in Eq. (12) is
\begin{equation}
	\textbf{v}_i(\tilde{t} +\delta t/2)= \textbf{v}_i(\tilde{t}-\delta t/2) +\frac{\delta t}{m_i} \textbf{f}_i(\tilde{t}).
\end{equation}
 The change in the discrete positions are obtained by
  inserting  (18) in Eq. (13) 
\begin{eqnarray}
	\textbf{r}_i(\tilde{t}+\delta t/2)=  \textbf{r}_i(\tilde{t})+ \frac{\delta t}{2} \textbf{v}_i(\tilde{t} +\delta t/2),
\end{eqnarray}
and the velocities and momenta are constant in between the changes at the discrete times so Eq. (20) is equivalent to
\begin{equation}
	\textbf{r}_i(\tilde{t}+\delta t)=  \textbf{r}_i(\tilde{t})+  \delta t\textbf{v}_i(\tilde{t}+\delta t/2),
\end{equation}
and Eqn. (19) and (21) are the Leapfrog formulation of Newton's discrete algorithm \cite{Toxvaerd1993}. 

MD simulations with Newton's discrete dynamics start   at time $t=0$ with two sets of start data:

$\textbf{r}_i(-\delta t)$ and  $\textbf{r}_i(0)$ (Verlet),

  $\textbf{r}_i(0)$ and  $\textbf{v}_i(-\delta t/2)=(\textbf{r}_i(0)$ - $\textbf{r}_i(-\delta t)/\delta t $

  (Leapfrog, velocity-Verlet),\\
  and if the position-Verlet algorithm is compared with the other algorithms with their  start data for
  the force actions at time zero  one shall start the position-Verlet algorithm with

$\textbf{r}_i(\tilde{t}=0)=\textbf{r}_i(\delta t/2)$ and $ \textbf{v}_i(\tilde{t}-\delta t/2)=  \textbf{v}_i(-\delta t/2)$. However, the simulations
in \cite{Ni2024} were started at $x_0=y_0 (\equiv \textbf{r}_i(-\delta t))$ and $x_1=y_1 (\equiv\textbf{r}_i(0))$ 
for both the velocity-Verlet and the position-Verlet algorithms \cite{comment1}.

\section{Newton's discrete dynamics}
Newton's discrete dynamics have the same qualities and invariances as his analytic dynamics. It is time reversible, symplectic, and with the
exact conservation of momentum, angular momentum, and energy for a conservative system. Newton's third law
ensures the two first invariances by which pairs of force actions between two objects cancel. The exact energy conservation, which is not obvious, can be seen by comparing the
work done by the forces in  time intervals  with the corresponding change in the kinetic energy. The proof is given in \cite{Toxvaerd2023,Toxvaerd2024} and in the Appendix.
 The  derivation of the discrete energy conservation is in fact in close analogy to the way energy conservation is
derived for Newton's analytic dynamics \cite{Goldstein}, and to the formulation of the first law of thermodynamics.

There is probably a remarkable connection between Newton's analytic and discrete dynamics: the existence of a
 $\textit{shadow Hamiltonian}, H_{shadow}$ nearby the Hamiltonian $H$ for the corresponding analytic dynamic.
 If the analytic dynamics with the shadow Hamiltonian  stars at the same phase point at $t=t_0$ as Newton's discrete dynamics,
 then the discrete points $\textbf{r}_i(t_n)$ at $t_0, t_0+\delta t,..,t_0+n\delta t),..$ are located on the
 analytic trajectory $\textbf{r}_i(t)$  for  $H_{shadow}$ \cite{Toxvaerd2023,Toxvaerd1994,Toxvaerd2012}.
 The existence of  $H_{shadow}$ means no qualitative differences exist 
 between the two kinds of dynamics. However, if one starts the force calculation 
 at another position corresponding to the position at a later time for the analytic dynamics then the discrete dynamics
 is with another nearby shadow Hamiltonian except for monotonic forces.
 The  linear extrapolation in the position-Verlet dynamics  from $\textbf{r}(t_n)$ to 
 $\textbf{r}(t_n+\delta t/2)$ before the forces are calculated is to  positions which are only on the trajectory of  $H_{shadow}$
 for the analytic dynamics from $t=t_0$  for
 forces which also depend linearly with positions. For all other force fields, the trajectories will be different. There are small numerical differences between
 positions obtained by different algorithms for Newton's discrete dynamics due to the accumulation of different round-off errors \cite{Rodger1989} which could be removed by
 performing the simulations with integer arithmetics \cite{Levesque1993}.

When discussing the qualities of different reformulations of Newton's discrete dynamics, one must
  compare the dynamics from the same
 starting point for force actions, as with analytic dynamics. The exact conservation of energy implies that the discrete dynamics
 are propagating on an energy shell in the microcanonical phase space.  The trajectories 
 for starting points with different force actions deviate,  and this is also closely analogous to what happens if one starts with different Hamiltonians for analytic dynamics.

Hundreds of thousands of articles with MD, and in all subdisciplines of Natural Sciences, have been published since Verlet published his MD simulations in 1967.
 Almost all the simulations are with Newton's discrete algorithm, and with the same qualities as Newton's analytic dynamics.  
 Feynman gave in 1982 a keynote speech \textit{Simulating Physics with Computers} \cite{Feynman1982} 
in which he talked  ``...about the possibility...that the computer will do
exactly the same as nature", and his conclusion was that it is not possible.
Newton's discrete dynamics is exact in the same sense as his analytic dynamics, but
 computer simulations are not exact simulations of  real systems dynamics, they contain many approximations.
 The physical world is not known exactly and it is far more complex than any simulated systems, and no real systems have been simulated exactly.
Hence, more than forty years later, and after hundreds of thousands of computer simulations of the physical system's dynamics
the answer to Feynman's question is still negative.
But although it is not possible to simulate the dynamics exactly for any real systems, 
simulations with Newton's discrete algorithm have been and will be of great use in Natural Science.

\appendix
\section{ The energy invariance in discrete Newtonian dynamics}
  Newton's classical discrete dynamics between $N$ spherically symmetrical objects
     with masses $ m^N=m_1, m_2,..m_i,..,m_N$ and positions \textbf{r}$^N(t)=$\textbf{r}$_1(t)$, \textbf{r}$_2(t)
     ,..,$\textbf{r}$_i(t),..$\textbf{r}$_N(t)$  is obtained 
 by Eq. (1). Let the force, $ \textbf{f}_i(t)$ on object No $i$ be a sum of pairwise  forces  $ \textbf{f}_{ij}(t)$ between pairs of   objects $i$ and $j$
 \begin{equation}
	 \textbf{f}_i(t)=  \sum_{j \neq i}^{N} \textbf{f}_{ij}(t).
 \end{equation}	
  Newton's algorithm is a  symmetrical time-centered difference whereby the dynamics is time reversible and symplectic.
  The conservation of momentum and angular momentum for a conservative system follows directly from Newton's third law  for the
  conservative system with $\textbf{f}_{ij}(t)=-\textbf{f}_{ji}(t)$, but the energy invariance is not so obvious.

 The  energy  in analytic dynamics is the sum of potential energy $ U(\textbf{r}^N(t))$ and kinetic energy $K(t)$, 
 and it is a time-invariance for a conservative  system.
 However, the kinetic energy  in the discrete dynamics  with a force action at time $t$ is not well-defined. 
Traditionally one uses Verlet's  first-order expression for the velocity at time $t$
\begin{eqnarray}
	\textbf{v}_{0,i}(t)=\frac{\textbf{v}_i(t+\delta/2)+\textbf{v}_i(t-\delta/2)}{2} \nonumber \\
	=\frac{\textbf{r}_i(t+\delta/2)-\textbf{r}_i(t-\delta/2)}{2 \delta t}
\end{eqnarray}
obtained by his time symmetric Taylor expansion \cite{Toxvaerd2023a}, and
 \begin{eqnarray}
	 K_0(t)=  \sum_i^N \frac{1}{2}m_i \textbf{v}_{0,i}(t)^2 \\
	E_0(t)= U(\textbf{r}^N(t))	+K_0(t)
 \end{eqnarray}	 
  The energy $E_0(t)$ obtained  by using  the approximations Eqn. (A3) and (A4) with $K(t)=K_0(t)$ for the kinetic energy and  $U(\textbf{r}^N(t))$ for
  the potential energy of analytic dynamics 
  fluctuates with time, although it is constant averaged over long time intervals \cite{comment2}.

  The velocities  in Newton's discrete dynamics are, however, constant in between the discrete times with force actions, and
the energy invariance can be obtained by dividing  time into  time intervals
 $[t_n- \delta t/2, t_n+ \delta t/2]$ with force actions at $t_n$ ($n$=1,2,...) and with 
sub-intervals $[t_n-\delta t/2,t_n]$ and $[t_n, t_n+\delta t/2,t]$.  
The energy invariance, $E_{\textrm{D}}$ in Newton's discrete dynamics (D)
 can then be  obtained by considering the change in kinetic energy $\delta K_ {\textrm{D}}$,
the work  $W_{\textrm{D}}$ done by the forces, and the change in discrete values of the  potential  energy   $\delta U_ {\textrm{D}} \equiv - W_{\textrm{D}}$
in  a time interval $[t-\delta t/2, t+\delta t/2]$ with $t=t_n$.

The loss in  potential energy, $-\delta U_{\textrm{D}}$ is defined as
the work done by the forces at a  move of the positions \cite{Goldstein}. 
The discrete force at time $t$   change
 the position  from  $(\textbf{r}_i(t)+ (\textbf{r}_i(t-\delta t))/2$ at $t-\delta t/2$
to   the position  $(\textbf{r}_i(t+\delta t)+ \textbf{r}_i(t))/2$ at $t+\delta t/2$,
and with the change $\delta \textbf{r}_i$ of the position
$ \delta \textbf{r}_i= \textbf(\textbf{r}_i(t+\delta t) -\textbf{r}_i(t-\delta t))/2$ and with 
\begin{eqnarray}
	-\delta U_{\textrm{D}} \equiv W_{\textrm{D}}= \sum_i^N  \textbf{f}_i(t)  \delta \textbf{r}_i \nonumber \\
 =\sum_i^N  \textbf{f}_i(t) (\textbf(\textbf{r}_i(t+\delta t) -\textbf{r}_i(t-\delta t))/2.
\end{eqnarray}	
By rewriting Eq. (4) to
\begin{equation}
	\textbf{r}_i(t+ \delta t) -\textbf{r}_i(t-\delta t)= 2(\textbf{r}_i(t) -\textbf{r}_i(t-\delta t))+\frac{\delta t^2}{m_i} \textbf{f}_i(t),
\end{equation}
and inserting in Eq. (A5) one obtains an expression for the total work in the time interval
\begin{equation}
	-\delta U_{\textrm{D}}=  W_{\textrm{D}}	=  \sum_i^N  [\textbf{f}_i(t)(\textbf(\textbf{r}_i(t) -\textbf{r}_i(t-\delta t)) + \frac{\delta t^2}{2m_i}\textbf{f}_i(t)^2].
\end{equation}

The mean kinetic energy $K_{\textrm{D}}$ of the discrete dynamics  in the time interval $[t-\delta t/2, t+\delta t/2]$ is

\begin{eqnarray}	
 K_{\textrm{D}}=                           \nonumber \\		
\frac{1}{2} \sum_i^N \frac{1}{2}m_i[\frac{\textbf(\textbf{r}_i(t+\delta t/2)-\textbf{r}_i(t))^2}{\delta (t/2)^2}+
\frac{\textbf(\textbf{r}_i(t)-\textbf{r}_i(t-\delta t/2))^2}{\delta (t/2)^2}] \nonumber \\
=\frac{1}{2}\sum_i^N\frac{1}{2}m_i [\frac{\textbf(\textbf{r}_i(t+\delta t)-\textbf{r}_i(t))^2}{\delta t^2}+
\frac{\textbf(\textbf{r}_i(t)-\textbf{r}_i(t-\delta t))^2}{\delta t^2}],
\end{eqnarray}
and with the change 
\begin{eqnarray}	
\delta K_{\textrm{D}}=                           \nonumber \\		
=\sum_i^N\frac{1}{2}m_i [\frac{\textbf(\textbf{r}_i(t+\delta t)-\textbf{r}_i(t))^2}{\delta t^2}-
\frac{\textbf(\textbf{r}_i(t)-\textbf{r}_i(t-\delta t))^2}{\delta t^2}].
\end{eqnarray},

By rewriting  Eq. (4) to
\begin{equation}
	\textbf{r}_i(t+ \delta t) -\textbf{r}_i(t)= \textbf{r}_i(t) -\textbf{r}_i(t-\delta t)+\frac{\delta t^2}{m_i} \textbf{f}_i(t)
\end{equation}
   and inserting the squared expression for  $\textbf{r}_i(t+\delta t) -\textbf{r}_i(t)$ in  Eq. (A9), the change in kinetic energy
   is
\begin{equation}
	\delta K_{\textrm{D}}= \sum_i^N [ \textbf{f}_i(t) (\textbf{r}_i(t)-\textbf{r}_i(t - \delta t)) +\frac{\delta t^2}{2m_i} \textbf{f}_i(t)^2].
\end{equation}
The energy invariance at a discrete change of time from $t-\delta t/2$ to  $t+\delta t/2$ 
in Newton's discrete dynamics is expressed by Eqn. (A7),  and  (A11) as \cite{Toxvaerd2023}
\begin{equation}
	\delta E_{\textrm{D}}=	\delta( U_{\textrm{D}}+ K_{\textrm{D}})=0.
\end{equation}

Unfortunately,
the energy $E_D$ in the MD simulations is recorded with systematic errors. The systematic errors are partly caused by the use of the analytical
expressions for the potential energies of the discrete forces\cite{comment2}, by truncating the potentials and not the forces \cite{Toxvaerd2011},
and partly by using the incorrect expression $K_0$ for the kinetic energy.
However, the errors are typically of a few percent or less \cite{Toxvaerd2024}.


\begin{thebibliography}{99}
\bibitem{comment1} The simulations with 	velocity-Verlet and position-Verlet in
\cite{Ni2024} are started from the same two sets of positions $x_0=y_0=0$ and $x_1=y_1=0.1=\tau/2$ (TABLE I in \cite{Ni2024}).
	If one compares the two algorithms at the same starting point in the phase space with $x_0=0 (\equiv \textbf{r}_i(-\delta t)),x_1=0.1 (\equiv \textbf{r}_i(0)) $
		for the start with the velocity-Verlet algorithm, then the position-Verlet shall be started not with  $y_0=0$ and $y_1=0.1=\tau/2
		( \equiv \delta t/2)$ but with $y_0=-\tau/2(\equiv \textbf{r}_i(-\delta t/2)) , y_1=\tau/2(\equiv \textbf{r}_i(\delta t/2))$.
\bibitem{comment2}  The potential energy between to objects, $i$ and $j$, at a distance $r_{ij}$, and  e.g. with a discrete Lennard-Jones (LJ)
	force is not the Lennard-Jones potential energy $u_{\textrm{LJ}}(r_{ij})$, but given by the total
		discrete work on an object by moving  it with discrete steps from infinity to $r_{ij}$,
 in analogy with the way the analytic energy is determined.
\end{thebibliography}
\end{document}